\documentclass[prd,letterpaper,twocolumn,preprintnumbers,nofootinbib]{revtex4}
\usepackage[letterpaper, hdivide={1.91cm,,1.165cm}, vdivide={1.83cm,,2.8cm}]{geometry}

\usepackage{amsmath,amssymb}
\usepackage{graphicx}
\usepackage[hyperfootnotes=false,colorlinks,citecolor=blue]{hyperref}

\newcommand{\dd}{\ensuremath{\text{d}}}
\newcommand{\ii}{\ensuremath{\text{i}}}

\allowdisplaybreaks

\begin{document}

\title{Q-balls across dimensions}

\author{Dusty Aiello}
\email[Email: ]{cty5qj@virginia.edu}
\thanks{ORCID: \href{https://orcid.org/0009-0003-4988-4548}{0009-0003-4988-4548}.}
\affiliation{Department of Physics, University of Virginia,
Charlottesville, Virginia 22904-4714, USA}

\author{Julian Heeck}
\email[Email: ]{heeck@virginia.edu}
\thanks{ORCID: \href{https://orcid.org/0000-0003-2653-5962}{0000-0003-2653-5962}.}
\affiliation{Department of Physics, University of Virginia,
Charlottesville, Virginia 22904-4714, USA}

\hypersetup{
pdftitle={Q-balls across dimensions},   
pdfauthor={Dusty Aiello, Julian Heeck}
}

\begin{abstract}
Scalars carrying a conserved global charge $Q$ can form stable localized field configurations composed of a large number of particles. These non-topological solitons are spherically symmetric and are called Q-balls. While usually analyzed in three spatial dimensions, these solitons can be straightforwardly generalized to $d$ spatial dimensions. For $d=1$, we can analytically solve the non-linear differential equation for an important class of single-field potentials; for $d>1$, we can analytically approximate the solutions in the thin-wall or large Q-ball regime, including the first sub-leading correction consistently. Since the underlying differential equations have the same form as vacuum-decay bounce solutions, our results find applications there, too.
\end{abstract}

\maketitle

\section{Introduction}
\label{sec:intro}

Classical scalar field theories can exhibit stable localized solutions. Derrick has shown that \textit{stationary} solutions of this kind are only possible for spatial dimensions $d <3$~\cite{Derrick:1964ww}. However, it is possible to construct time-dependent field solutions that \textit{appear} stationary  even in $d=3$ as long as the theory contains some conserved global charge $Q$, for example due to a global $U(1)$ symmetry of the Lagrangian~\cite{Rosen:1968mfz,Friedberg:1976me}. The best-known example of such solitons are Coleman's Q-balls~\cite{Coleman:1985ki}, which can be realized in certain $U(1)$-symmetric complex-scalar potentials involving attractive-force terms. The resulting objects become the lowest-energy configuration with charge $Q$ for sufficiently large $Q$, and can be interpreted as stable macroscopic bound states of $Q$ scalars.

Most work on Q-balls has been focused on three spatial dimensions, $d=3$, matching our universe, but the construction straightforwardly generalizes to arbitrary integer dimensions $d\geq 1$. This is, by and large, more mathematics than physics, but at least the cases $d=4$ and $d=3$ happen to be relevant beyond Q-balls: the underlying differential equations plus boundary conditions match those of \textit{vacuum-decay bounce solutions} -- which was Coleman's starting point historically -- in  vacuum~\cite{Coleman:1977py} and in a thermal background~\cite{Linde:1981zj}, respectively.

We set out to study Q-balls in $d$ spatial dimensions in a class of single-field potentials that allow for good analytical approximations while being flexible enough to be useful stand-ins for other potentials that can accommodate Coleman-like Q-balls, identified in Ref.~\cite{Heeck:2022iky}.
Mathematically, we set this up by taking one (classical) complex field $\phi(x^\mu)$, which behaves according to the Klein--Gordon equation of motion: 
\begin{align}
    \left(\frac{\partial^2}{\partial t^2}-\vec{\nabla}^2\right)\phi = -\frac{\partial U}{\partial\phi^*},
    \label{eq:KleinGordon}
\end{align}
where $\vec{\nabla}$ is the $d$-dimensional nabla operator and $U$ is the potential that describes self-interactions of the scalar field. We are interested in localized solutions, so $\phi$ should vanish at spatial infinity. While time-independent localized solutions are impossible according to Derrick's theorem~\cite{Derrick:1964ww} -- at least for $d\geq 3$ --  solutions in which only the $\phi$ phase changes with time, $\phi\propto e^{\ii \omega t}$, exist and lead to outwardly time-independent objects called Q-balls, as shown by Coleman. The quantities of interest are the Q-ball energy $E$ (i.e.~its mass) and its charge $Q$ (i.e.~the number of scalars inside),
\begin{align}
    {E}&=\int{}^{}\dd^dx\left[\frac{\partial}{\partial t}{\phi}\frac{\partial}{\partial t}{\phi^*}+{\vec{\nabla}\phi}{\vec{\nabla}\phi^*+U}\right], \\ {Q}&=-\ii\int{}^{}\dd^dx\left[{\phi^*}\frac{\partial}{\partial t}{\phi}-{\phi}\frac{\partial}{\partial t}{\phi^*}\right],
\end{align}
which are constant for the field configurations under investigation here.

We are of course not the first to investigate Q-balls in $d$ dimensions: 
Refs.~\cite{Gleiser:2005iq,Tsumagari:2008bv} use  variational approaches to study the $d$-dependence for $d\geq 2$, which correspond to the leading term in Coleman's thin-wall expansion. We will systematically include sub-leading corrections and also study the case $d=1$, which is actually the simplest one, allowing for exact analytic solutions.

The rest of this article is organized as follows: we motivate and define the class of potentials of interest in Sec.~\ref{sec:model}, also setting up the differential equation and macroscopic Q-ball parameters. Sec.~\ref{sec:1d} contains the analytic solution for the $d=1$ case. For $d>1$, analytic solutions in our class of potentials are impossible, and we outline how to obtain numerical solutions in Sec.~\ref{sec:numerics}. Sec.~\ref{sec:thinwall} contains analytic approximations of the $d>1$ Q-balls in the thin-wall limit, including sub-leading effects. Finally, we conclude in Sec.~\ref{sec:conclusion}. Appendix~\ref{app:derivations} contains derivations of known results such as $\dd E/\dd \omega = \omega \dd Q/\dd \omega$ and the virial theorem for the convenience of the reader.

\section{Model}
\label{sec:model}

The scalar potential $U$ should be a function of $\phi^*\phi$ with global minimum at $\phi=0$ to ensure a conserved $U(1)$ charge, but also contain negative terms that correspond to attractive self-interactions. The simplest class of potentials satisfying these requirements are polynomials of the form
\begin{align}
U(|\phi|) = m^2_\phi |\phi|^2 -\beta |\phi|^p+\xi |\phi|^q \,,
\label{eq:phi_potential}
\end{align}
with $2 < p < q$ and positive constants $m_\phi^2$, $\beta$ and $\xi$~\cite{Heeck:2022iky}. In general spatial dimension $d$, such a potential is not renormalizable, but can always be interpreted as an effective field theory~\cite{Heeck:2022iky}. Exceptions arise for $d=1$, where scalar fields are dimensionless and \textit{any} integer $p$ and $q$ are renormalizable, and $d=2$, where $(p,q)=(4,6)$ is perturbatively renormalizable. To keep our analysis manageable, we will restrict ourselves to the case of \textit{equidistant exponents}, $p-2=q-p \equiv n$, i.e.~$p=2+n$, $q=2+2n$ with positive $n$, which allows for some analytical considerations and is an excellent approximation for many Coleman-type potentials, esp.~if we allow for non-integer $n$~\cite{Heeck:2022iky}. Well-known cases for $d=3$ include $n=1$~\cite{Kusenko:1997ad} and $n=2$~\cite{Heeck:2020bau}.

\begin{figure}[tb]
    \centering
    \includegraphics[width=0.48\textwidth]{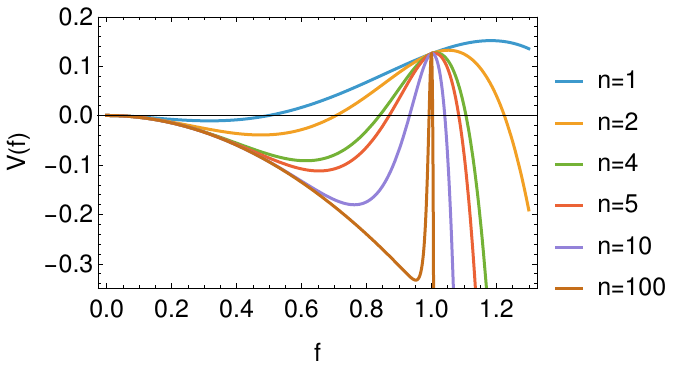}
    \caption{Effective potential $V(f)$ from Eq.~\eqref{eq:potential} for  $\kappa =1/2$.}
    \label{fig:potential}
\end{figure}

Coleman showed that Q-ball solutions require $U(|\phi|)/|\phi|^2$ to have a minimum at $|\phi|\equiv \phi_0/\sqrt2$ such that
\begin{align}
0 \leq \sqrt{\frac{ U(\phi_0/\sqrt2)}{(\phi_0/\sqrt2)^2}} \equiv \omega_0 < m_\phi \,.
\label{eq:minimum}
\end{align}
The case $\omega_0=0$ is special since $U$ then has two degenerate minima~\cite{PaccettiCorreia:2001wtt}.
In our potential~\eqref{eq:phi_potential}, we find
\begin{align}
    \phi_0 = \sqrt2 \left(\frac{\beta}{2\xi}\right)^{1/n}\,, &&
    \omega_0 = \sqrt{ m_\phi^2 - \frac{\beta^2}{4\xi}} \,.
\end{align}
The lowest-energy localized field configuration with non-zero $Q$ is spherically symmetric, i.e.~a ball, and takes the form $\phi=e^{\ii\omega t}f(r)\,\phi_0/\sqrt2$~\cite{Coleman:1985ki}, with dimensionless radial profile $f(r)$ and chemical potential $\omega \in (\omega_0 , m_\phi)$ that satisfies $\omega = \dd E/\dd Q$~\cite{Lee:1991ax} (see App.~\ref{app:derivations}).
We define the dimensionless quantities that replace $\omega$ and the radial coordinate $r$,
\begin{align}
    \kappa^2 \equiv\frac{\omega^2-\omega_0^2}
{m_\phi^2-\omega_0^2} \,, &&
\rho \equiv r\sqrt{m_\phi^2-\omega_0^2} \,,
\label{eq:kappa_definition}
\end{align}
to rewrite the Klein--Gordon equation as a simple \textit{one-parameter} differential equation
\begin{align}
    f''(\rho) +\frac{d-1}{\rho} f'(\rho) + \frac{\dd}{\dd f} V(f) = 0\,,
    \label{eq:eom}
\end{align}
with effective potential (illustrated in Fig.~\ref{fig:potential})
\begin{align}
    V(f) = -\frac{1}{2} f^2 \left( 1-\kappa^2  - 2 f^n + f^{2n}\right)
    \label{eq:potential}
\end{align}
and boundary conditions $f'(0)=0$ and $f(\rho\rightarrow\infty)=0$. We search for monotonically decreasing non-negative solutions $f(\rho)$, which correspond to the Q-ball ground states~\cite{Volkov:2002aj,Mai:2012cx,Almumin:2021gax}.

For fixed $n$, the potential depends only on the dimensionless variable $\kappa$, which is defined to lie between 0 and~1~\cite{Heeck:2020bau}.
The differential equation~\eqref{eq:eom} can be interpreted as a classical-mechanics problem in which a particle rolls in the potential $V$ subject to a friction force $(d-1)f'/\rho$, with $\rho$ and $f$ playing the roles of time and particle position, respectively~\cite{Coleman:1985ki}. According to the boundary conditions, the particle starts at rest from some position $f(0)$ and rolls to the local maximum at $f=0$, where it comes to rest after an infinitely long time. Since this journey starts and ends with zero kinetic energy, we need $V(f(0)) \geq 0$, the exact point being determined by the energy lost to friction.

Of particular interest is the case of small $\kappa$, where the global maximum of $V$ at 
\begin{align}
f_+ &\equiv \left(\frac{2+n+\sqrt{n^2 + 4 \kappa^2 + 4 n \kappa^2}}{2+2n}\right)^{\frac{1}{n}}\\
&= 1 + \frac{\kappa^2}{n^2} - \frac{(1+3n)\kappa^4}{2n^4} + \mathcal{O}(\kappa^6)\,.
\label{eq:fplus}
\end{align}
becomes degenerate with the local maximum at $f=0$:
\begin{align}
V(f_+)= \frac{\kappa^2}{2} +\frac{\kappa^4}{2n^2} + \mathcal{O}(\kappa^6)\,.
\label{eq:Vfplus}
\end{align}
The particle can then sit approximately force free near $f_+$ until the time-dependent friction term has eventually died off, and then swiftly roll from $f_+\simeq 1$ to $f=0$. The small-$\kappa$ solution therefore resembles a step-function that allows for analytical expressions for $E$ and $Q$ that have been used to prove the stability of these large thin-wall Q-balls~\cite{Coleman:1985ki}. As shown in Refs.~\cite{Heeck:2020bau,Heeck:2022iky}, the class of potentials from Eq.~\eqref{eq:potential} allows for a simple analytical improvement over Coleman's step function, vastly improving our understanding of thin-wall Q-balls.

The integrals for $E$ and $Q$ take the following form in our rescaled variables (see App.~\ref{app:derivations}):
\begin{align}
    {Q}&= \frac{2\pi^{d/2}\omega \phi_0^2}{\Gamma (\frac{d}{2})(m_\phi^2-\omega_0^2)^{d/2}}\int_0^\infty\dd \rho\, \rho^{d-1}f^2\,, \label{eq:finalQintegral}\\
    E&=\omega Q+\frac{2\pi^{d/2}\phi_0^2}{d\,\Gamma (\frac{d}{2}) (m_\phi^2-\omega_0^2)^{(d-2)/2}}\int_0^\infty\dd  \rho\, \rho^{d-1}f'^2 \,,\label{eq:finalEintegral}
\end{align}
and we furthermore define the (dimensionless) Q-ball radius $R$ via $f''(R)=0$. For fixed $d$ and $n$, the two integrals and the radius only depend on $\kappa$. The two integrals are not independent though: the well-known relation $\dd E/\dd\omega = \omega \dd Q/\dd \omega$ translates into
\begin{align}
    \int_0^\infty\dd \rho\,\rho^{d-1}f^2&=-\frac{1}{d\, \kappa} \frac{\dd}{\dd\kappa}\left[\int_0^\infty\dd \rho\,\rho^{d-1}(f')^2\right]
     \,,
    \label{eq:integral_relations}
\end{align}
so the $\kappa$ dependence of one integral is sufficient to determine the other (see App.~\ref{app:derivations}).
Of particular interest is furthermore the ratio $E/(m_\phi Q)$, the ratio of Q-ball energy to the energy of $Q$ scalars at rest. For $E < m_\phi Q$, the Q-ball is stable, which is typically the more interesting scenario. In our notation, and using Eq.~\eqref{eq:integral_relations},
\begin{align}
    \frac{E}{m_\phi Q}&=\frac{\omega}{m_\phi}+\frac{1}{d}\left(\frac{m_\phi^2-\omega_0^2}{m_\phi\omega}\right)\frac{\int_0^\infty\dd \rho\,\rho^{d-1}(f')^2}{\int_0^\infty\dd \rho\,\rho^{d-1}f^2}\label{eq:EoverQm} \\
    &=\frac{\omega}{m_\phi}+\left(\frac{m_\phi^2-\omega_0^2}{2m_\phi\omega}\right)\left( \frac{\dd \log\int_0^\infty\dd \rho\,\rho^{d-1}(f')^2}{\dd\kappa^2}\right)^{-1} ,\nonumber
\end{align}
which also depends explicitly on $\omega_0$.

Our goal is to solve, or at least approximate, the differential equation~\eqref{eq:eom} for various $d$ and $n$ to learn something about the macroscopic Q-ball quantities $Q$, $E$, and $R$.

Our results are not just of interest to Q-ball enthusiasts, but also find applications in the study of vacuum decay~\cite{Coleman:1977py}, where $f$ plays the role of the \textit{bounce} that interpolates between real and false vacuum. At zero temperature, the bounce equation is of the form Eq.~\eqref{eq:eom} with $d=4$; for thermal tunneling, one needs $d=3$, assuming a universe with three spatial dimensions. Our analysis and class of potentials can then also be employed to model bounce solutions. The quantity of interest there is the Euclidean bounce action, which is proportional to $E-\omega Q$ in Q-ball notation, or the integral  $\int_0^\infty\dd \rho\,\rho^{d-1}(f')^2$~\cite{Espinosa:2023osv}.

\section{Exactly solving \texorpdfstring{$d=1$}{d=1}}
\label{sec:1d}

While exact solutions often prove difficult or even impossible to find, our differential equation~\eqref{eq:eom} has a convenient property for $d=1$: the friction term vanishes and the equation can be integrated once analytically to find the conserved mechanical energy $\mathcal{E} \equiv \frac{1}{2}f'^2+V$. Given our boundary conditions, $\mathcal{E}=0$, allowing us to formally write the solution as
\begin{align}
\rho (f) = \int_{0}^{\rho}\dd\tilde\rho =\int_{f(0)}^{f}\frac{\dd \tilde f}{\sqrt{-2V(\tilde f)}} \,.
\end{align}
Conservation of $\mathcal{E}$ means $f(0)$ needs to have $V=0$, specifically the zero to the left of the global maximum (see Fig.~\ref{fig:potential}), which gives $f(0) = (1-\kappa)^{1/n}$. The above integral can then be performed analytically and solved for $f(\rho)$, a neat feature of these equidistant polynomial potentials:
\begin{align}
    f(\rho)=\left(\frac{1-\kappa^2}{1+\kappa \cosh\left[n\,\sqrt{1-\kappa^2} \,\rho\right]}\right)^{1/n} .
\end{align}
The Q-ball radius, from $f''(R)=0$, takes the form:
\begin{align}
    R=\frac{\cosh^{-1}\left(\frac{n + \sqrt{n^2 + 4(n+1)\kappa^2}}{2\kappa}\right)}{n\sqrt{1-\kappa^2}} \,,\label{eq:REqn_d1}
\end{align}
which diverges for $\kappa \to 0$ and $\kappa\to 1$:
\begin{align}
    R\to \begin{cases}
        \log (2n/\kappa)/n \,, \text{ for } \kappa\to 0\,,\\
        \frac{\cosh^{-1}\left(1+n\right)}{\sqrt2 \, n\, \sqrt{1-\kappa}}\,, \text{ for } \kappa\to 1\,.
    \end{cases}
\end{align}
$f(\rho)$ looks like a Gaussian for $\kappa\sim \mathcal{O}(1)$, but exhibits step-function behavior for $\kappa\ll 1$, as expected from Coleman's thin-wall argument; for $\kappa \to 1$, $f$ becomes more diffuse as $f(0)\to 0$ and $R\to \infty$. 

The two integrals that determine $E$ and $Q$ can be solved in terms of hypergeometric and beta functions,
\begin{align}
 \int{}{}\dd \rho \, f^2 &=   \frac{(1-\kappa)^{\frac{2}{n}} B\left(\tfrac{1}{2}, \tfrac{2}{n}\right)\,
{}_2F_1 \left(\tfrac{1}{2}, \tfrac{2}{n}; \tfrac{4+n}{2n}; \frac{1-\kappa}{1+\kappa}\right)}
{n \sqrt{1-\kappa^2}}\,,\\
\begin{split}
    \int{}{}\dd \rho\, f'^2 &=\frac{4 \kappa^{2}}{n}
(1+\kappa)^{-\frac{2(n+1)}{n}} 
(1-\kappa^{2})^{\frac{1}{2}+\frac{2}{n}} 
B \left(\tfrac{3}{2}, \tfrac{2}{n}\right) \\
&\quad\times {}_2F_1 \left(\tfrac{3}{2}, \tfrac{2+2n}{n}; \tfrac{4+3n}{2n}; \tfrac{1-\kappa}{1+\kappa}\right),
\end{split}
\end{align}
which satisfy the identity of Eq.~\eqref{eq:integral_relations}. For $n=1$ and $n=2$, these can be written in terms of logarithms and other more elementary functions.
The $f'$ integral is monotonically decreasing from $n/(4+2n)$ at $\kappa=0$ to $0$ at $\kappa=1$; the $f$ integral diverges near $\kappa =0$ as $-\log(\kappa)/n$, but the behavior near $\kappa=1$ depends critically on $n$: for $n< 4$ ($n>4$), the integral goes to zero (diverges) for $\kappa\to 1$. $n=4$ is special in that the $f$ integral goes to the finite value $\pi/\sqrt{32}$.
This also implies that $Q$ diverges for $\kappa\to 1$ if $n>4$ and vanishes for $n <4$, even though the Q-ball radius diverges in all cases.

With the two integrals at our disposal, we can study the stability of $d=1$ Q-balls through the ratio $E/(m_\phi Q)$ from Eq.~\eqref{eq:EoverQm}. For fixed $\kappa$ and $\omega_0$, $E/(m_\phi Q)$ grows with $n$ and eventually becomes bigger than 1 since $\lim_{n\to \infty} E/(m_\phi Q) = m_\phi/\omega > 1$. 
For the special case $\omega_0 = 0$, we find that all Q-balls are unstable for $n \geq 4$; for $n < 4$, $E/(m_\phi Q)$ is smaller than 1 for $\kappa \in (0.075,1)$ ($n=1$),  $\kappa \in (0.30,1)$ ($n=2$),  $\kappa \in (0.62,1)$ ($n=3$), so stable Q-balls have a maximal charge $Q$ beyond which they become unstable.

\begin{figure*}[tbh]
    \centering
    \includegraphics[width=0.5\textwidth]{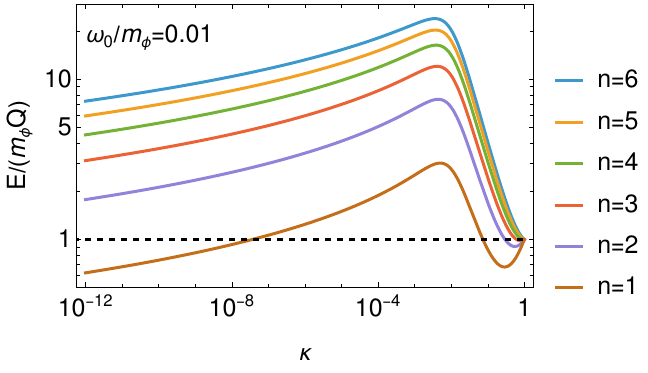}
    \includegraphics[width=0.42\textwidth]{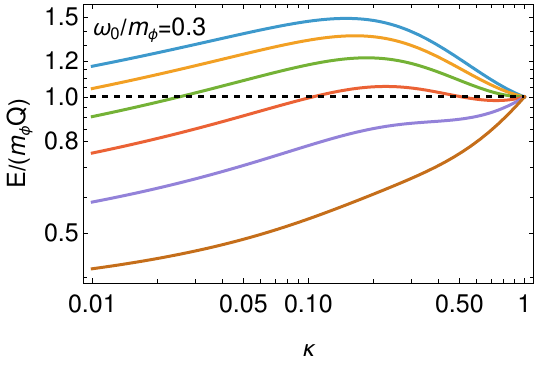}
    \caption{$E/(m_\phi Q)$ for $d=1$ as a function of $\kappa$ for $\omega_0/m_\phi = 0.01$ (left) and $0.3$ (right). Different colors correspond to different $n$, and stable Q-balls are in the region $E/(m_\phi Q) < 1$, i.e.~below the black dashed line. }
    \label{fig:EoverQ_d1}
\end{figure*}

For $0<\omega_0<m_\phi$, the behavior changes dramatically, and the Q-balls become stable for any fixed $n$ in the thin-wall small-$\kappa$ regime, quantitatively
\begin{align}
    \frac{E}{m_\phi Q} = \frac{\omega_0}{m_\phi} + \frac{n}{2+n} \frac{\phi_0^2 \sqrt{m_\phi^2 - \omega_0^2}}{m_\phi Q} + \mathcal{O}(e^{-Q}) \,,
    \label{eq:stability_d1}
\end{align}
with large $Q\propto - \log(\kappa)/n$.
For $\kappa\to 1$, $E/(m_\phi Q)$ goes to $1$, but the derivative is proportional to $n-4$, so Q-balls near $\kappa \sim 1$ are stable for $n< 4$ (but $Q\to 0$, so these are not large Q-balls) and unstable for $n>4$; for $n=4$, Q-balls are stable near $\kappa\sim 1$ for $\omega_0/m_\phi \geq \sqrt{5/8}\simeq 0.79$ and unstable otherwise. Overall, stable large-$Q$ solitons require $\omega_0\neq 0$ and $\kappa \ll 1$, without restrictions on $n$. 
For $\kappa$ away from the end points 0 and 1, i.e.~for not large Q-balls, the stability depends critically on $\omega_0$ and can feature transitions from stable to unstable and back as a function of $\kappa$, see Fig.~\ref{fig:EoverQ_d1} for an illustration.

\begin{figure}[tbh]
    \centering
    \includegraphics[width=0.42\textwidth]{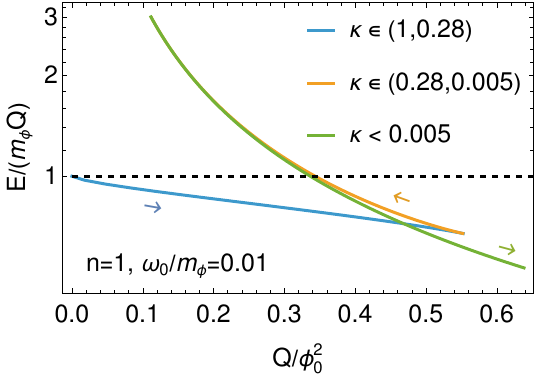}
    \caption{$E/(m_\phi Q)$ for $d=n=1$ as a function of $Q$ for $\omega_0/m_\phi = 0.01$. Different colors correspond to different $\kappa$ regions and the arrow indicate flow of decreasing $\kappa$. }
    \label{fig:EoverQ_Q_d1}
\end{figure}

The minima and maxima of $E/(m_\phi Q)$ as a function of $\kappa$ (Fig.~\ref{fig:EoverQ_d1}) correspond to turning points  when viewed as functions of $Q$. For example, the $n=1$ case with $\omega_0/m_\phi = 0.01$ features a minimum of $E/(m_\phi Q)$ at $\kappa \simeq 0.28$ and a maximum at $\kappa\simeq 0.005$. As a function of $Q$ (Fig.~\ref{fig:EoverQ_Q_d1}), this implies a decreasing $E/(m_\phi Q)$ vs.~$Q$ as $\kappa$ decreases from $1$ to $0.28$ (blue line in Fig.~\ref{fig:EoverQ_Q_d1}), followed by an increase until $\kappa\simeq 0.005$ (orange line), and then again decreasing $E/(m_\phi Q)$ (green line), corresponding to the large-$Q$ expression of Eq.~\eqref{eq:stability_d1}. Even though there is a stable configuration for every $Q$, it is not clear how Q-ball growth would look like in this region.

This concludes our discussion of Q-balls in $d=1$ spatial dimensions, the only analytically tractable scenario.

\section{Numerical solutions for \texorpdfstring{$d>1$}{d>1}}
\label{sec:numerics}

For $d>1$, the differential equation~\eqref{eq:eom} cannot be solved analytically, and is even nontrivial to solve numerically given the non-linear nature and the boundary condition at infinity. One way to solve it is the  \textit{shooting method}~\cite{Coleman:1985ki}, which consists of making an initial educated guess for $f(0)$, and then manually varying $f(0)$ to ensure $f\to 0$ at infinity. While straightforward, this method quickly becomes tedious for large Q-balls.

A more convenient approach is to first map the infinite domain of $\rho$ onto a finite domain by applying the coordinate transformation
\begin{align}
    y\equiv \frac{\rho}{1+\frac{\rho}{a}}\,,
    \label{eq:Coord_transformation}
\end{align}
 where $a$ is some large number, say $a=1000$~\cite{Heeck:2020bau}.
The coordinate $y$ is now in the finite range between 0 and $a$, and our differential equation for $f(y)$ takes the form
\begin{align}
    \left(1-\frac{y}{a}\right)^4\left(\frac{\dd^2f}{\dd y^2}+\frac{\dd f}{\dd y}\frac{1}{y}\cdot\frac{\left(d-1\right)-2\frac{y}{a}}{1-\frac{y}{a}}\right)+\frac{\dd V}{\dd f}=0\,,
\end{align}
with boundary conditions $f'(0) = 0 = f(a)$. 
This differential equation can now be solved using \texttt{Mathematica}'s Finite-Element Method, which requires some guess function $f(y)$ as input and should converge to the true solution if the functions are close to each other. For small Q-balls, this approach is fast and does not require particularly tuned guess functions; for large Q-balls, the problem becomes more stiff and either requires excellent guess functions (for example the analytic approximations derived below) or a crawling loop in which $\kappa$ is reduced by small increments $\Delta \kappa$, using the solution $f_{\kappa}$ as the guess function for $f_{\kappa - \Delta \kappa}$. For very small $\kappa$, thin-wall approximations allow analytic treatment, perfectly complementing the numerical solutions.

Finally, the vacuum-decay solver \texttt{Anybubble}~\cite{Masoumi:2017trx} supports arbitrary spacetime dimensions and works well and fast outside of the large Q-ball regime. We compared many of our solutions to \texttt{Anybubble} and found excellent agreement. Data files for $d=2$--$6$ with $n=1$--$6$ are available as ancillary files of the arXiv version of this article~\cite{Aiello:2026jay}.

\section{Analytic approximations for \texorpdfstring{$d>1$}{d>1}}
\label{sec:thinwall}

\begin{figure*}[tbh]
    \centering
    \includegraphics[width=0.49\textwidth]{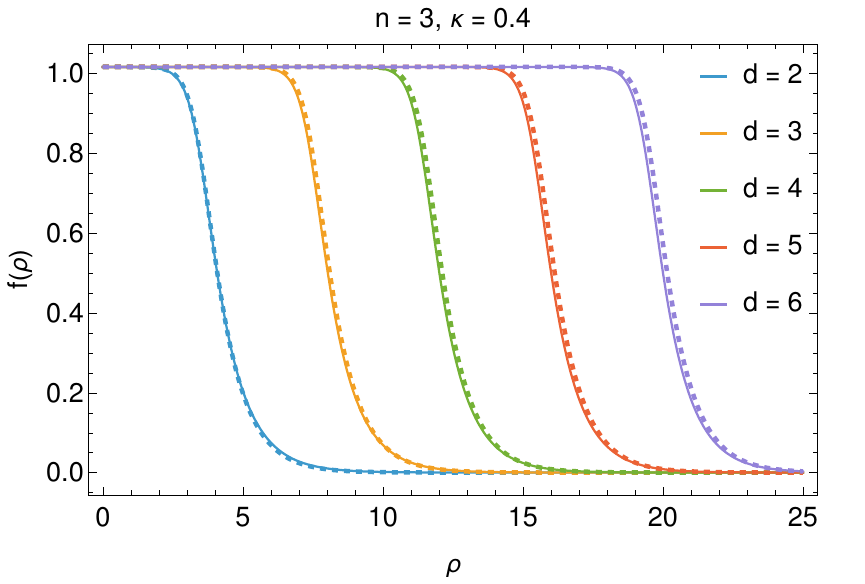}
    \includegraphics[width=0.48\textwidth]{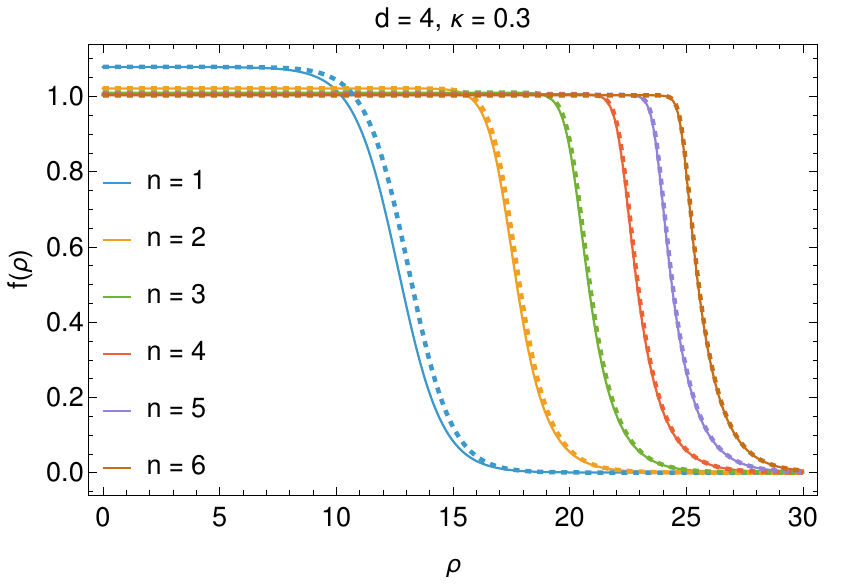}
    \caption{Q-ball profiles $f(\rho)$ for fixed $n=3$ and several $d$ (left) and fixed $d=4$ and several $n$ (right). The solid lines are the numerical solutions, the dashed lines our analytical approximations from Eq.~\eqref{eq:f_new} with radius from Eq.~\eqref{eq:radius}. Notice that we picked rather large $\kappa$ in order to see at least some deviations.}
    \label{fig:profiles}
\end{figure*}

\begin{figure*}[tbh]
    \centering
    \includegraphics[width=0.48\textwidth]{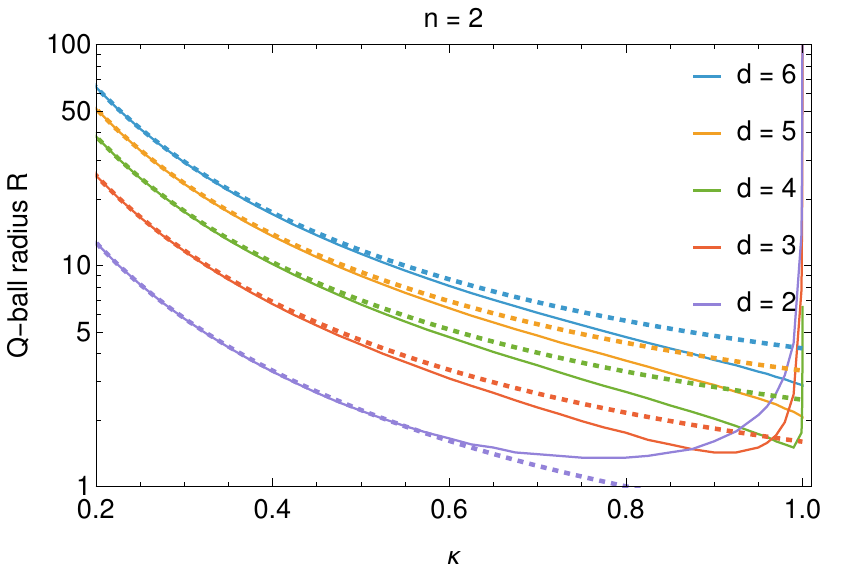}
    \includegraphics[width=0.48\textwidth]{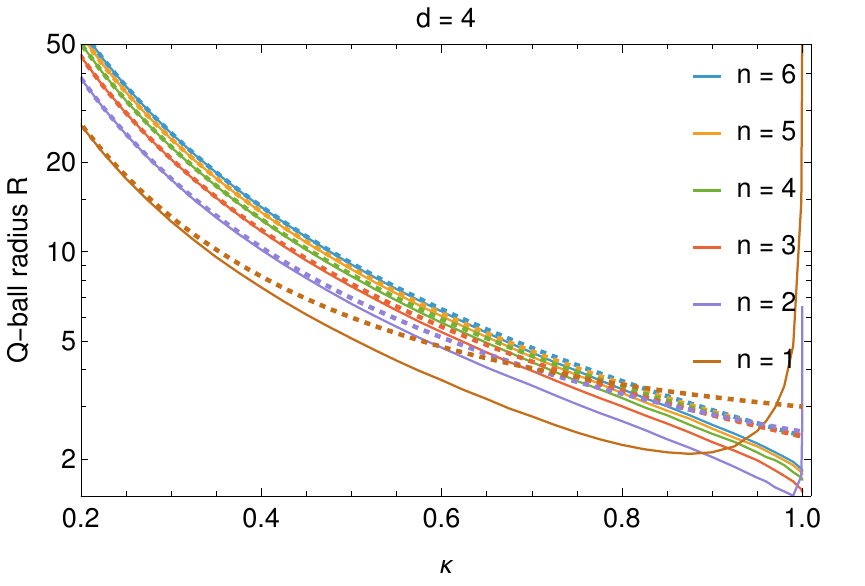}
    \caption{
    Q-ball radii for fixed $n=2$ and several $d$ (left) and fixed $d=4$ and several $n$ (right). The solid lines are the numerical solutions, the dashed lines our analytical approximations from Eq.~\eqref{eq:radius}.}
    \label{fig:radius}
\end{figure*}

The best known and most important approximation for $d>1$ Q-balls arises in the thin-wall limit, $\kappa \ll 1$, in which $f(\rho)$ resembles a step function that drops from 1 to 0 at some radius $\rho = R$, which can be obtained as in Refs.~\cite{Heeck:2020bau,Heeck:2022iky} to be 
\begin{align}
    R = \frac{2n}{2 + n} \cdot \frac{(d - 1)}{2} \cdot \frac{1}{\kappa^2} 
    \label{eq:radius_old}
\end{align}
in the limit $\kappa\to 0$.
As expected, small $\kappa$ correspond to large Q-ball radii; notice the faster growth compared to $d=1$, where $R\propto \log \kappa$. Following Ref.~\cite{Heeck:2022iky}, we can also obtain $f(\rho)$ in the limit $\kappa\to 0$, resulting in the transition function that improves upon the simple step-function approximation, essentially smoothing it out:
\begin{align}
    f_0(\rho) = \frac{1}{\left( 1 + n e^{n (\rho - R)} \right)^{1/n}} \,,
    \label{eq:profile_old}
\end{align}
where the radius is defined via $f''(R)=0$. Eqs.~\eqref{eq:radius_old} and~\eqref{eq:profile_old} can also be viewed as the leading terms in the proper small-$\kappa$ expansion of the solution to Eq.~\eqref{eq:eom}, recently employed to the vacuum-decay case in Refs.~\cite{Ivanov:2022osf,Matteini:2024xvg}. In that approach, we shift our coordinate to $\rho = R + z$ and expand both profile and radius as power series in $\kappa^2$, using Eqs.~\eqref{eq:radius_old} and~\eqref{eq:profile_old} as starting points~\cite{Heeck:2020bau}:
\begin{align}
    f(z) &= f_0 (z) + \kappa^2 f_1(z)+\kappa^4 f_2(z)+\ldots\,,\\
    R &= \frac{R_0}{\kappa^2} + R_1 + \kappa^2 R_2+\ldots\,.
\end{align}
The differential equation~\eqref{eq:eom} can then be solved order by order in $\kappa^2$ after the friction term is expanded:
\begin{align}
    f'' + V'(f)\Big|_{\kappa=0}+\kappa^2 f &= - \frac{d-1}{R+z}f'\\
    &=- \frac{\kappa^2 (d-1) f'}{R_0} \\
    &\quad + \frac{\kappa^4 (d-1) (z+R_1) f'}{R_0^2}+\ldots\nonumber
\end{align}
The boundary conditions are now $f'(z=-R\to -\infty)$ and $f(z\to\infty)=0$. The leading-order term $f_0$ indeed just gives back~\eqref{eq:profile_old},  while the order $\kappa^2$ equation takes the form
\begin{align}
    \left(\frac{\dd^2}{\dd z^2}+ V''(f_0)\Big|_{\kappa=0}\right) f_1 = -\frac{d-1}{R_0} f_0' - f_0\,.
    \label{eq:f1EOM}
\end{align}
The leading-order solution $f_0$ is translation invariant, i.e.~the radius is not constrained; equivalently, $f_0'(z)$ is a zero mode of the differential operator $\dd^2/\dd z^2+ V''(f_0)|_{\kappa=0}$. According to Fredholm, solving for $f_1$ is then only possible if the source term is orthogonal to the zero mode:
\begin{align}
    \int_{-\infty}^\infty \dd z \, f_0'(z) \left[-\frac{d-1}{R_0} f_0'(z) - f_0(z)\right] = 0\,.
\end{align}
Performing these integrals leads to the $R_0$ equation
\begin{align}
    R_0 = \frac{2n}{2 + n} \cdot \frac{(d - 1)}{2} \,,
    \label{eq:R0}
\end{align}
corresponding exactly to the leading term from Eq.~\eqref{eq:radius_old}.
We had already obtained $f_0$ and $R_0$ using other arguments, but the method at hand allows us to obtain subleading corrections. With $R_0$ fixed by Eq.~\eqref{eq:R0}, we can solve Eq.~\eqref{eq:f1EOM} to obtain $f_1(z)$:
\begin{align}
    f_1 (z) = \frac{1}{n^2} f_0(z) + \frac{z}{n} f_0'(z)\,.
\end{align}
The $\mathcal{O}(\kappa^4)$ equation for $f_2$ takes the form
\begin{align}
    \left(\frac{\dd^2}{\dd z^2}+ V''(f_0)\Big|_{\kappa=0}\right) f_2 &= -\frac{d-1}{R_0} f_1' +\frac{d-1}{R_0^2} (z+R_1)f_0' \nonumber\\
    &\quad - f_1- \frac{1}{2} f_1^2 V'''(f_0)\Big|_{\kappa=0}\,.
\end{align}
Demanding again that the source term on the right-hand side is orthogonal to the zero mode $f_0'$ fixes $R_1$, which after a lengthy calculation comes out to
\begin{align}
    R_1=\frac{d-2 +  \gamma + \log n + \psi^{(0)}\left(\frac{2}{n}\right) +\frac{1-d}{n+2}}{n} \,,
\end{align}
with the Digamma function $\psi^{(0)} (x)\equiv \Gamma'(x)/\Gamma(x)$ and Euler--Mascheroni constant $\gamma\simeq 0.577$.
We will not attempt to find $f_2$ or $R_2$, but make one additional modification; to order $\kappa^2$, the initial position $f(\rho=0)$ is $1+\kappa^2/n^2$. We will replace this by the location of the global $V$ maximum, $f_+$, which includes additional subleading $\mathcal{O}(\kappa^4)$ terms~\cite{Heeck:2020bau}, see Eq.~\eqref{eq:fplus}. Our final thin-wall approximation is therefore
\begin{align}
    f(\rho) &= f_+ f_0(\rho-R) + \frac{\kappa^2}{n} (\rho-R) f_0'(\rho-R)
    \label{eq:f_new}
\end{align}
with radius
\begin{align}
    R &= \frac{n(d-1)}{(2 + n)\kappa^2} 
    + \frac{d-2 +  \gamma + \log n + \psi^{(0)}\left(\frac{2}{n}\right) +\frac{1-d}{n+2}}{n} \,,
    \label{eq:radius}
\end{align}
This is an excellent approximation for small $\kappa$, see for example Fig.~\ref{fig:profiles}. Notice that dropping the $f_0'$ part in $f$ would be fine too, it is the radius equation that is more important, and Eq.~\eqref{eq:radius} provides an excellent approximation even for not-so-small $\kappa$, as illustrated in Fig.~\ref{fig:radius}.

We can now use Eq.~\eqref{eq:f_new} to calculate the two integrals that determine $Q$ and $E$. Expanding out the squares in $f^2$ and $f'^2$ and using integration by parts, we need three base integrals, which can be written in terms of hypergeometric functions:
\begin{widetext}
\begin{align}
    \int_0^\infty\dd \rho\,\rho^{d-1}f_0^2 &= 2^{-d}\, n^{-2/n}\, \Gamma(d)\, e^{2R}\,
{}_{{d+1}}F_{{d}}\!\left(
\underbrace{\frac{2}{n},\ldots,\frac{2}{n}}_{d+1\ \text{times}};
\underbrace{\frac{n+2}{n},\ldots,\frac{n+2}{n}}_{d\ \text{times}};
-\frac{e^{nR}}{n}
\right) \\
&=\frac{R^d}{d} 
    - \frac{1}{n} \left( \gamma + \log n + \psi^{(0)}\left(\frac{2}{n}\right) \right) R^{d - 1}  + \mathcal{O}\left(R^{d-2}\right) ,
    \label{eq:intF} \\
    %
\int_0^\infty\dd \rho\,\rho^{d-1}(f_0')^2 &=  2^{-d}\, n^{-2/n}\, \Gamma(d)\, e^{2R}\,
{}_{{d+1}}F_{{d}}\!\left(
\underbrace{\frac{2}{n},\ldots,\frac{2}{n}}_{d\ \text{times}},\,
\frac{2n+2}{n};
\underbrace{\frac{n+2}{n},\ldots,\frac{n+2}{n}}_{d\ \text{times}};
-\frac{e^{nR}}{n}
\right)\\
&=\frac{n R^{d - 1}}{4 + 2n}  
    - \frac{(d - 1)\left(\gamma-1 + \log n + \psi^{(0)}\left(\frac{2}{n}\right)\right)}{2(2 + n)} R^{d - 2} + \mathcal{O}\left(R^{d-3}\right) ,
     \label{eq:intFP} \\
    %
\int_0^\infty\dd \rho\,\rho^{d-1}(f_0'')^2 &=  2^{-d}\, n^{-2/n}\, \Gamma(d)\, e^{2R}\,
{}_{{d+1}}F_{{d}}\!\left(
\underbrace{\frac{2}{n},\ldots,\frac{2}{n}}_{d\ \text{times}},\,
\frac{4n+2}{n};
\underbrace{\frac{n+2}{n},\ldots,\frac{n+2}{n}}_{d\ \text{times}};
-\frac{e^{nR}}{n}
\right)\nonumber\\
&\quad -  \frac{2 \Gamma(d)\, e^{(2+n)R}}{n^{2/n}\,(2+n)^{d}}\,\,
{}_{{d+1}}F_{{d}}\!\left(
\underbrace{\frac{n+2}{n},\ldots,\frac{n+2}{n}}_{d\ \text{times}},\,
\frac{4n+2}{n};
\underbrace{\frac{2n+2}{n},\ldots,\frac{2n+2}{n}}_{d\ \text{times}};
-\frac{e^{nR}}{n}
\right) \\
&\quad +  \frac{\Gamma(d)\, e^{(2+2n)R}}{2^{d}\,  n^{2/n}\,(1+n)^{d}}\,\,
{}_{{d+1}}F_{{d}}\!\left(
\underbrace{\frac{2n+2}{n},\ldots,\frac{2n+2}{n}}_{d\ \text{times}},\,
\frac{4n+2}{n};
\underbrace{\frac{3n+2}{n},\ldots,\frac{3n+2}{n}}_{d\ \text{times}};
-\frac{e^{nR}}{n}
\right) \nonumber\\
&=\frac{n^3 R^{d - 1}}{8+16n+6n^2}  
    - \frac{(d - 1)n^2\left(n-3 + 2\log n +2\gamma+ 2\psi^{(0)}\left(\frac{2}{n}\right)\right)}{4(2+n)(2+3n)} R^{d - 2} + \mathcal{O}\left(R^{d-3}\right) .
    \label{eq:intFPP} 
\end{align}
\end{widetext}

\begin{figure*}[tbh]
    \centering
    \includegraphics[width=0.48\textwidth]{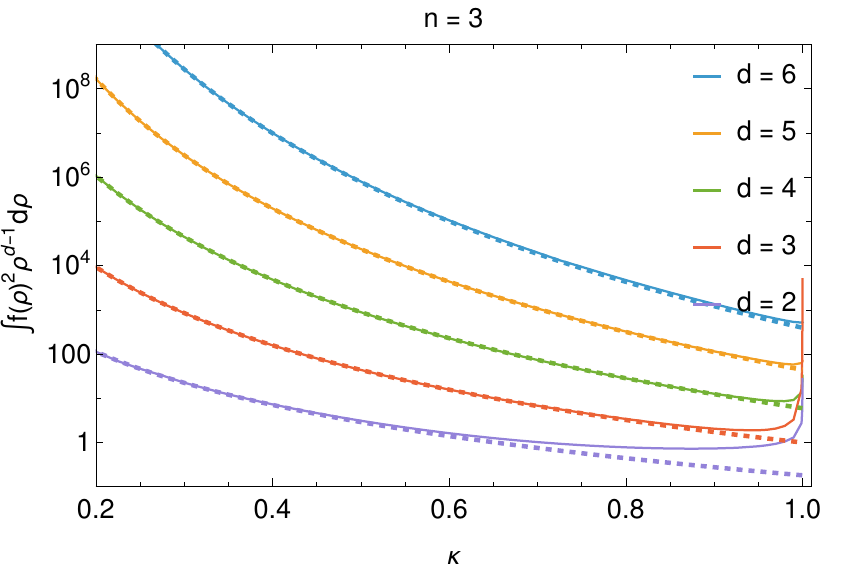}
    \includegraphics[width=0.48\textwidth]{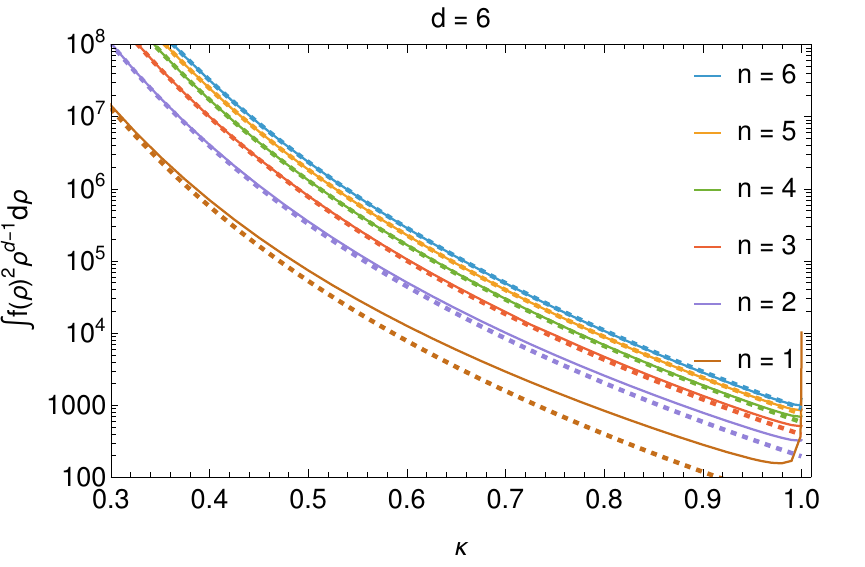}
    \caption{
    Q-ball integrals relevant for mass and charge for fixed $n=3$ and several $d$ (left) and fixed $d=6$ and several $n$ (right). The solid lines are the numerical solutions, the dashed lines our analytical approximations from Eq.~\eqref{eq:final_intf}.}
    \label{fig:intf}
\end{figure*}

\begin{figure*}[tbh]
    \centering
    \includegraphics[width=0.48\textwidth]{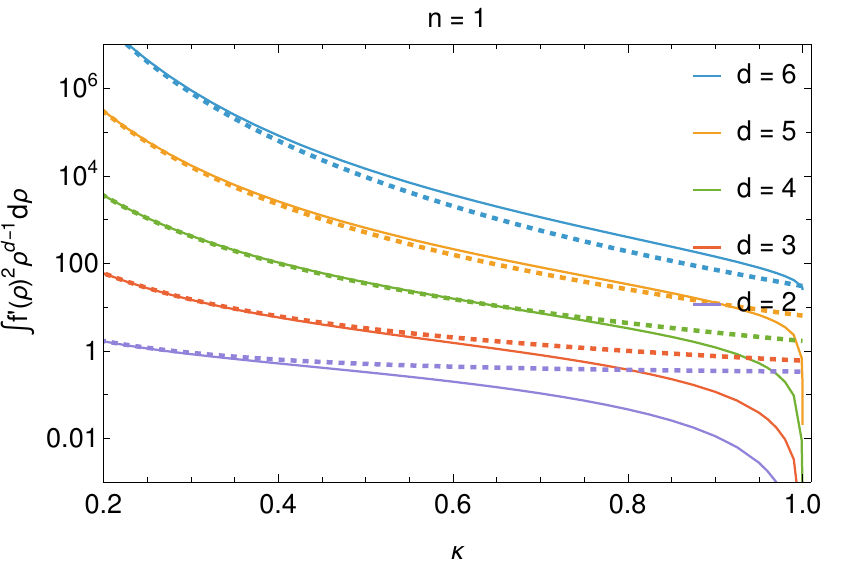}
    \includegraphics[width=0.48\textwidth]{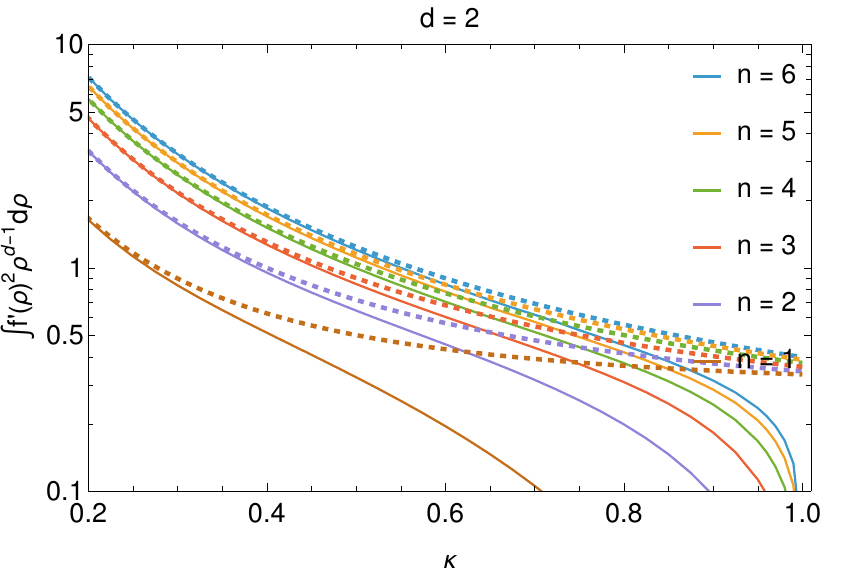}
    \caption{
    Q-ball integrals relevant for mass for fixed $n=1$ and several $d$ (left) and fixed $d=2$ and several $n$ (right). The solid lines are the numerical solutions, the dashed lines our analytical approximations from Eq.~\eqref{eq:final_intfp}.}
    \label{fig:intfp}
\end{figure*}

With these integrals at our disposal and our expression for $f$ from Eq.~\eqref{eq:f_new}, we eventually find the rather compact expressions
\begin{align}
    \int\dd \rho\,\rho^{d-1}f^2 &= \frac{(d-1)^d n^d}{d (n+2)^d \kappa^{2d}}\nonumber \\
    &\quad\times \left( 1+ \frac{(d-2)(1+d+n d)}{(d-1)n^2}\kappa^2\right) , \label{eq:final_intf}\\
    \int\dd \rho\,\rho^{d-1}f'^2 &= \frac{(d-1)^{d-1} n^d}{2 (n+2)^d \kappa^{2d-2}} \left( 1+ \frac{1+d+n d}{n^2}\kappa^2\right),\label{eq:final_intfp}
\end{align}
which satisfy Eq.~\eqref{eq:integral_relations} to first sub-leading order. Notice that all the Digamma pieces cancel thanks to our consistent thin-wall expansion, unlike previous expressions that effectively employed partial results~\cite{Heeck:2020bau,Heeck:2022iky}; the leading terms agree, of course.
These two expressions agree very well with the numerical solutions, see Figs.~\ref{fig:intf} and~\ref{fig:intfp}, and become exact for $\kappa\to 0$, where both integrals diverge.
Using the leading terms in the stability ratio~\eqref{eq:EoverQm},
\begin{align}
 \frac{E}{m_\phi Q} &\simeq \frac{\omega_0}{m_\phi} +  \frac{d}{d-1}\,\frac{1-\omega_0^2/m_\phi^2}{2\,\omega_0/m_\phi} \kappa^2\\
&\simeq \frac{\omega_0}{m_\phi} +  \frac{2^{1/d} d n \sqrt{\pi}
\,
\sqrt{1-\frac{\omega_0^2}{m_\phi^2}}}
{2(2+n)\,(d\,\Gamma(d/2))^{1/d} \omega_0}
\left(
\frac{\phi_0^2 \omega_0}{Q}
\right)^{1/d} ,
\end{align}
we see immediately that the small-$\kappa$ regime leads to stable Q-balls for all $n$ and $d>1$,\footnote{For $\omega_0=0$, we  find $E/m_\phi Q \simeq (1+1/(2d-2))\kappa \propto Q^{-1/(2d-1)}$.} generalizing results that were obtained previously at fixed $n$~\cite{Gleiser:2005iq}. The $Q^{-1/d}$ scaling of the sub-leading term was also derived in Ref.~\cite{Gleiser:2005iq}.
Amusingly, taking $d\to 1$ in the last equation, $E/m_\phi Q$ as a function of $Q$, gives the same result we found in the $d=1$ case, see Eq.~\eqref{eq:stability_d1}, which also holds to next order since the $Q^{-2/d}$ coefficient vanishes for $d=1$.

\section{Conclusion}
\label{sec:conclusion}

We have investigated Q-balls in arbitrary spatial dimensions $d$ for a family of potentials that contains many popular models but is flexible enough to be a good stand-in for most Coleman-like single-field potentials. The case $d=1$ can be solved analytically and is qualitatively different from $d>1$, although it still features a stable-Q-ball thin-wall regime.
For $d>1$, we have to resort to analytic approximations, for which we constructed a systematic thin-wall expansion. We consistently include the first correction to the typical thin-wall transition profile and $R\propto 1/\kappa^2$ radius relationship, leading to compact approximations for the macroscopic Q-ball quantities of radius, energy, and charge.
Given the structural equivalence between Q-ball equations and vacuum decay, we hope our results can also be of use as analytic approximations for bounce solutions.


\section*{Acknowledgments}
We thank Mikheil Sokhashvili for discussions.
This work was supported by the U.S.~Department of Energy under Grant No.~DE-SC0007974. We acknowledge Research Computing at The University of Virginia for providing computational resources that have contributed to the results reported within this publication.
Numerical data files for $d=2$--$6$ with $n=1$--$6$ are available as ancillary files of the arXiv version of this article~\cite{Aiello:2026jay}.


\appendix

\section{Derivations}
\label{app:derivations}

For the convenience of the reader, we provide a few derivations about $d$-dimensional Q-ball energy and charge that have been used in the main text. These results are well known~\cite{Zhou:2024mea} and depend on the space dimension $d$ in a fairly obvious way, but might be of interest to the uninitiated.

\subsection{Energy--charge relationship}

First off, we show $\frac{\dd E}{\dd {\omega}}=\omega\frac{\dd Q}{\dd\omega}$, starting from the  definitions of $Q$ and $E$:
\begin{align}
    {Q}&=-\ii\int{}^{}\dd^dx\left[{\phi^*}\frac{\partial}{\partial t}{\phi}-{\phi}\frac{\partial}{\partial t}{\phi^*}\right],\\
    {E}&=\int{}^{}\dd^dx\left[\frac{\partial}{\partial t}{\phi}\frac{\partial}{\partial t}{\phi^*}+{\vec{\nabla}\phi}{\vec{\nabla}\phi^*+U}\right] .
\end{align}
Using Coleman's Q-ball ansatz,
$\phi(x)=\frac{\phi_0}{\sqrt2}f(r)e^{\ii\omega t}$, these expressions simplify to
\begin{align}
    {Q}&=\phi_0^2\omega\int{}^{}\dd^dxf^2 \,, \label{eq:Qfintegral}\\
    {E}& =\phi_0^2\int{}^{}\dd^dx\left[\frac{\omega^2}{2}f^2(r)+\frac{1}{2}f'^2(r)+\frac{U(f)}{\phi_0^2}\right]. \label{eq:Efintegral}
\end{align}
where $f'$ is the derivative with respect to the radial coordinate $r$. Next, we write $E=\omega Q $ plus a remainder:
\begin{align}
    {E}&=\omega Q+\phi_0^2\int{}^{}\dd^dx\left[-\frac{\omega^2}{2}f^2(r)+\frac{1}{2}f'^2(r)+\frac{U(f)}{\phi_0^2}\right] ,\label{eq:E_Definition}
\end{align}
and then take the $\omega$ derivative:
\begin{align}
    \frac{\dd E}{\dd\omega}=\omega\frac{\dd Q}{\dd\omega}+\phi_0^2\int{}^{}\dd^dx
    \left[f'\frac{\dd f'}{\dd\omega}-\omega^2f\frac{\dd f}{\dd\omega}+\frac{1}{\phi_0^2}\frac{\dd U}{\dd\omega}\right] .\label{eq:dE/dw_simpl}
\end{align}
We use the Klein--Gordon equation~\eqref{eq:KleinGordon} and our ansatz for $\phi$ to find 
\begin{align}
    \left(\left(\ii\omega\right)^2-\vec{\nabla}^2\right)f=-\frac{\dd U}{\dd f} \,.\label{eq:Heeck_fDiffEq}
\end{align}
The next observation is that $U$ depends on $\omega$ only through $f$; therefore, we are able to use the chain rule and write $\frac{\dd U}{\dd\omega}=\frac{\dd U}{\dd f}\frac{\dd f}{\dd\omega}$. With this and equation~\eqref{eq:Heeck_fDiffEq} we can rewrite equation~\eqref{eq:dE/dw_simpl} as
\begin{align}
    \frac{\dd E}{\dd\omega}=\omega\frac{\dd Q}{\dd\omega}+&{\frac{2\pi^{\frac{d}{2}}}{\Gamma\left(\frac{d}{2}\right)}}\phi_0^2\int{}^{}\dd r\,r^{d-1}
    \left[f'\frac{\dd f'}{\dd\omega}\right. \nonumber\\
    &\left.-\omega^2f\frac{\dd f}{\dd\omega}+(\omega^2f+\vec{\nabla}^2f)\frac{\dd f}{\dd\omega}\right],
\end{align}
also performing the $d$-dimensional angular integrals. 
The terms involving $\omega^2$ inside the integral cancel and we are left with an integral containing only $f$ and its derivatives. The Laplace operator in $d$-dimensional spherical coordinates for a function that only depends on $r$ is $\Delta f(r) = \frac{1}{r^{d-1}}\frac{\partial}{\partial r} \left( r^{d-1} \frac{\partial f(r)}{\partial r}\right)$, so we arrive at
\begin{align}
    \frac{\dd E}{\dd\omega}=\omega\frac{\dd Q}{\dd\omega}+&{\frac{2\pi^{\frac{d}{2}}}{\Gamma\left(\frac{d}{2}\right)}}\phi_0^2\int{}^{}\dd r\, r^{d-1}
    \left[f'\frac{\dd f'}{\dd\omega}\right.\nonumber\\
    &\left.+\frac{1}{r^{d-1}}\frac{\partial}{\partial r} \left( r^{d-1} \frac{\partial f(r)}{\partial r}\right)\frac{\dd f}{\dd\omega}\right].
\end{align}
By performing integration by parts on the second term in the integrand we observe some convenient cancellations and end up with
\begin{align}
   \frac{\dd E}{\dd\omega}=\omega\frac{\dd Q}{\dd\omega}+{\frac{2\pi^{\frac{d}{2}}}{\Gamma\left(\frac{d}{2}\right)}}\phi_0^2\left[r^{d-1}f'\frac{\dd f}{\dd\omega}\right]_{r=0}^{r=\infty} \,.
\end{align}
The term that remains from partial integration vanishes due to the fact that $f'(0)=0$ and $f'(\infty)=0$, according to our boundary conditions. We are left with the desired relation
\begin{align}
    \frac{\dd E}{\dd\omega}=\omega\frac{\dd Q}{\dd\omega}.
    \label{eq:dEdw}
\end{align}
Up to some special points, this also implies $\dd E/\dd Q = \omega$, allowing us to identify $\omega$ with the chemical potential, i.e.~the change in energy due to a change in the number of particles.

\subsection{Virial theorem}

Next, we derive the virial theorem, or Derrick's theorem~\cite{Derrick:1964ww}, allowing us to relate average potential and kinetic energy, and ultimately rewrite our expression for $E$ without an integral over $U$. Following a process similar to that of Gleiser and Thorarinson~\cite{Gleiser:2005iq}, we start with Eqs.~\eqref{eq:Qfintegral} and~\eqref{eq:Efintegral} and rewrite the $\omega^2$ term in $E$ in terms of $Q^2$:
\begin{align}
    E&=\phi_0^2\int{}^{}\dd^dx\left[\frac{1}{2}f'^2(r)+\frac{U(f)}{\phi_0^2}\right]+\frac{Q^2}{2\phi_0^2\int{}^{}\dd^dxf^2}\,.
\end{align}
We  perform a rescaling of variables from $x\rightarrow\alpha x$, and hence  $\dd^dx\rightarrow\alpha^d\dd^dx$ and $f'\rightarrow\frac{1}{\alpha}f'$. The energy  becomes
\begin{align}
    E=\phi_0^2\int{}^{}\alpha^d\dd^dx\left[\frac{1}{2}\frac{1}{\alpha^2}f'^2+\frac{U(f)}{\phi_0^2}\right]+\frac{1}{\alpha^d}\frac{Q^2}{\phi_0^2\int{}^{}\dd^dxf^2} \,.
\end{align}
For fixed $Q$, the Q-ball solution should be the field configuration with the lowest possible energy according to Coleman, and every other $f(r)$ with the same $Q$ will have a larger energy. In particular, $f(\alpha r)$ will have a larger energy than $f(r)$ unless $\alpha=1$. We thus know that $E(\alpha)$ must have a minimum at $\alpha=1$, leading us to  $\frac{\partial E}{\partial\alpha}|_{\alpha=1}=0$:
\begin{align}
   0&=  \left.\frac{\partial E}{\partial\alpha}\right|_{\alpha=1}\\
   &=\phi_0^2\int{}^{}\dd^dx\left[\frac{1}{2}\left(d-2\right)\alpha^{d-3}f'^2+d\alpha^{d-1}\frac{U(f)}{\phi_0^2}\right]_{\alpha=1}\nonumber\\
   &\quad-\left[d\alpha^{-d-1}\frac{Q^2}{2\phi_0^2\int{}^{}\dd^dxf^2}\right]_{\alpha=1}\\
    &=\phi_0^2\int{}^{}\dd^dx\left[\frac{1}{2}\left(d-2\right)f'^2+d\frac{U(f)}{\phi_0^2}\right]-d\frac{Q^2}{2\phi_0^2\int{}^{}\dd^dxf^2}.
\end{align}
During the rescaling, $Q$ does not change, so we can return the expression to the notation involving $\omega^2$:
\begin{align}
    0=\phi_0^2\int{}^{}\dd^dx\left[\frac{1}{2}\left(d-2\right)f'^2+d\frac{U(f)}{\phi_0^2}-\frac{d}{2}\omega^2f^2\right] .
\end{align}
From here, we clearly see the relationship between the average potential energy and the average kinetic energy, the virial theorem:
\begin{align}
    d\int{}{}\dd^dx\left[\frac{U(f)}{\phi_0^2}\right]=\int{}{}\dd^dx\left[\frac{2-d}{2}f'^2+\frac{d}{2}\omega^2f^2\right] .\label{eq:VirialThm}
\end{align}
Using this expression, we can eliminate the integral over $U$ in the definition of $E$ and obtain the remarkably simple expression
\begin{align}
    E&=Q\omega+\phi_0^2\int{}{}\dd^dx\left[\frac{1}{d}f'^2\right] \\&=Q\omega+\frac{2\pi^{\frac{d}{2}}}{d\,\Gamma (\frac{d}{2})}\phi_0^2\int{}{}\dd r\,r^{d-1}f'^2\,,
\end{align}
performing the $d$-dimensional angular integrals in the last step. 
This is a convenient expression of the Q-ball energy, both practically and conceptually. Practically, energy and charge are determined by two simple integrals, one over $f^2$ and one over ${f'}^2$, see Eqs.~\eqref{eq:finalQintegral} and~\eqref{eq:finalEintegral}, which are far easier than integrating the potentially complicated function $U$. Conceptually, we find that the energy boils down to number of scalars ($Q$) times the chemical potential ($\omega$), plus a term that ends up being subleading in the limit of large Q-balls. In complete analogy to nuclear physics, we can call these terms volume and surface term, respectively. By using this expression together with $ \frac{\dd}{\dd\omega}\left(E-\omega Q\right)=-Q$ from Eq.~\eqref{eq:dEdw}, we can get a differential relation between the integrals over $f^2$ and ${f'}^2$:
\begin{align}
    \frac{\dd}{\dd\omega}\left[\frac{1}{d}\int{}{}\dd r\,r^{d-1}f'^2\right]&=-\omega\int{}{}\dd r\,r^{d-1}f^2 \,.
\end{align}
Switching from $\omega$ to the $\kappa$ defined in Eq.~\eqref{eq:kappa_definition}, we obtain Eq.~\eqref{eq:integral_relations} in the main text. 
This relationship proves that all Q-ball properties are determined by one integral, or better yet one integral as a function of $\omega$. The other integral follows from the above equation, and Q-ball energy and charge follow from the expressions above. Once we have obtained the radial Q-ball profile $f$ by solving the differential equation, we then just have to calculate one of these integrals.

\bibliographystyle{utcaps_mod}
\bibliography{BIB.bib}

\end{document}